\documentclass[12pt,JChemPhys]{JArticle}
\usepackage{lettercount}
\usepackage{mathmacros}

\def\titre{Two-component plasma in a gravitational field:
Thermodynamics}

\begin{document}
\title{\titre}
\author{Gabriel T\'ellez\thanks{e-mail address: gtellez@physique.ens-lyon.fr}\\%
{\rm \small \it Laboratoire de Physique}%
\thanks{Laboratoire associ\'e au Centre National de la Recherche
Scientifique - URA 1325} \\
{\rm\small\it Ecole Normale Sup\'erieure de Lyon,
46 All\'ee d'Italie, 69364 Lyon cedex 07, France}
}
\date{}
\maketitle

\begin{abstract}
We revisit the model of the two-component plasma in a gravitational
field, which mimics charged colloidal suspensions. We concentrate on the
computation of the grand potential of the system. Also, a special sum
rule for this model is presented. 
\end{abstract}

\section{Introduction}

In a recent paper,\citenote{tcp+g} the author presented a particular
solvable model of Coulomb system inspired from the problem of the
sedimentation equilibrium of charged colloidal suspensions.
Although the model is much simpler than real colloidal suspensions, it
features several properties observed in numerical results for more
realistic models\citenote{BibHans} and shows some new interesting
results. 

This Communication is a complement to Ref.~\cite{tcp+g} in which only
the density profiles where computed. Here we concentrate on
thermodynamic quantities such as the grand potential.  In
section~\ref{model} we briefly describe the model, in
section~\ref{GrandPotential} we compute the grand potential of the
system, and in the last section we present a special sum rule for this
model.

\section{The model}\label{model}

We consider a two-dimensional two-component plasma composed of two
species of particles with charges $\pm q$ and masses $M_\pm$. We
choose a system of Cartesian coordinates $(Oxy)$ in which the
gravitational field is $\mathbf{g}=-g\hat{\mathbf{y}}$. The particles
are in a container of height $h$ and infinite width. Let us define the
inverse gravitational lengths of the particles $k_\pm = \beta M_\pm
g$, $k_0=(k_+ + k_-)/2$, and $\dk=k_+ - k_-$, where $\beta$ is the
inverse temperature.  In the grand canonical ensemble we define
position-dependent rescaled fugacities $m(\r)=m_0 \exp (-k_\pm y)$ to
account for the interaction of the particles with the external
gravitational field. The screening length (at zero altitude) is given
by $1/m_0$.\citenote{tcp+g,CorJan}
	
The two-component plasma is equivalent to a free Dirac
field\citenote{CorJan,Coleman,Samuel} when $\beta q^2=2$. The grand
partition function can be written as $\Xi=\det(1+K)$, where
\begin{equation}
K=\left[m_+(\r)\frac{1+\sigma_z}{2}
+m_-(\r)\frac{1-\sigma_z}{2}\right]
\dslash^{-1}\,,
\end{equation}
with $\sigma_x$, $\sigma_y$, $\sigma_z$ the Pauli matrices and
$\dslash=\sigma_x \partial_x + \sigma_y \partial_y$.
To compute $\Xi$, one must solve the eigenvalue problem
\begin{equation}\label{Kcouples}
K\Psi=\lambda\Psi\,,
\end{equation}
where $\Psi=(\psi,\chi)$ and $\lambda$ are the eigenvectors and
eigenvalues of $K$. Then the grand potential is given by 
\begin{equation}\label{Omega-general}
\Omega=-k_B T \sum_\lambda \ln (1+\lambda)\,.
\end{equation}

When an external field is acting differently over the positive and
negative particles, as in the present case, it is useful to write the
fugacities as $m_\pm(\r)=m(\r)\exp[-(\pm)2V(\r)]$ where
$m(\r)=m_0\exp[-k_0 y]$ and $V(\r)=\dk\,y/4$. Let us 
define the auxiliary eigenfunctions
\begin{eqnarray}
\phi_+(\r)&=&e^{V(r)}\psi(\r)\,,\\
\phi_-(\r)&=&e^{-V(r)}\chi(\r)\,,
\end{eqnarray}
and the operators
\begin{equation}
a=\partial_{x}+i\partial_y+\partial_x V(\r)+i\partial_y V(\r)\,,
\end{equation}
and
\begin{equation}
a^{\dag}=-\partial_{x}+i\partial_y+\partial_x V(\r)-i\partial_y V(\r)\,.
\end{equation}
Then the eigenvalue problem~(\ref{Kcouples}) is equivalent to
\begin{equation}\label{Kdecouples}
\phi_+(\r)+\lambda^2 a^{\dag}m(\r)^{-1}a\left[m(\r)^{-1}\phi_+(\r)\right]=0
\,,
\end{equation}
\begin{equation}
\phi_-(\r)=\lambda a \left[m(\r)^{-1}\phi_+(\r)\right]\,,
\end{equation}
and the boundary conditions that $\psi$ ($\chi$) on the boundary is
equal to a function which is analytic (anti-analytic) outside the
container and vanishes at infinity.\citenote{JanManPis,JancoTellez1}
This method is general and can be applied to others models of
two-component plasma in an external field.

\section{The grand potential}\label{GrandPotential}

In the present case, we look for solutions of~(\ref{Kdecouples}) of
the form $\phi_+(\r)=h(y)\exp(ikx-k_0 y)$. From~(\ref{Kdecouples}) we
find the equation for $h(y)$
\begin{equation}
\left[(k+\dk/4)k_0-(k+\dk/4)^2-\lambda^{-2}m_0^2e^{-2k_0y}\right]
h(y)+k_0 h'(y)+h''(y)
=0\,,
\end{equation}
which is very similar to the equation~(3.2) of Ref.~\cite{tcp+g}
satisfied by the Green function $\hat g_{++}$. The solution is
\begin{equation}
h(y)=\left[AI_\nu\left(\frac{m_0e^{-k_0 y}}{\lambda k_0}\right)
+BK_\nu\left(\frac{m_0e^{-k_0 y}}{\lambda k_0}\right)
\right]e^{-k_0y/2}\,,
\end{equation}
with $\nu=|k-(k_-/2)|/k_0$, and $A$ and $B$ constants of
integration.  

In the present case, the boundary conditions become
$\phi_+(x,y=0)=0$ and $\phi_-(x,y=h)=0$ if $k>0$, and
$\phi_+(x,y=h)=0$ and $\phi_-(x,y=0)=0$ if $k<0$. These conditions give
a system of two linear homogeneous equations for the constants $A$ and
$B$. Writing that the discriminant of this system must be zero, in
order to have non-trivial solutions, gives the equation that determines
$\lambda$. Let $\sigma$ be the sign of $k-(k_-/2)$. If $k>0$,
$\lambda$ is a solution of
\begin{equation}
I_\nu\left(\frac{m_0}{\lambda k_0}\right)
K_{\nu+\sigma}\left(\frac{m_0e^{-k_0 h}}{\lambda k_0}\right)
+
I_{\nu+\sigma}\left(\frac{m_0e^{-k_0 h}}{\lambda k_0}\right)
K_\nu\left(\frac{m_0}{\lambda k_0}\right)
=0\,,
\end{equation}
and if $k<0$, $\lambda$ is given by
\begin{equation}
\label{eq:lambdakneg}
I_\nu\left(\frac{m_0e^{-k_0 h}}{\lambda k_0}\right)
K_{\nu-1}\left(\frac{m_0}{\lambda k_0}\right)
+
I_{\nu-1}\left(\frac{m_0}{\lambda k_0}\right)
K_\nu\left(\frac{m_0e^{-k_0 h}}{\lambda k_0}\right)
=0\,.
\end{equation}
These equations have an infinite number of solutions. For each $k$, we
may index the different values of $\lambda$ by an integer $\ell$:
$\lambda=\lambda_{k,\ell}$. Then, the grand potential by unit length
in the $x$-direction is
\begin{equation}
\omega=-\frac{k_B T}{2\pi}\int \sum_\ell \ln (1+\lambda_{k,\ell})\,dk \,.
\end{equation}
The sum over $\ell$ can be done
explicitly,\citenote{JanManPis,JancoTellez1,Forrester} for example in
the case where $k<0$, by noting that the zeros of the entire function
\begin{eqnarray}
f(z)=\moko e^{-k_0h\nu}z
\Big[
\lefteqn{
I_\nu(m_0e^{-k_0 h}z/k_0)
K_{\nu-1}(m_0 z/k_0)
}
\nonumber\\
&+
I_{\nu-1}(m_0 z/k_0)
K_\nu(m_0e^{-k_0 h}z/k_0)
\Big]\,,
\end{eqnarray}
are $1/\lambda_{k,\ell}$ and that $f(0)=1$, so $f(z)=\prod\limits_\ell
(1-\lambda_{k,\ell}z)$. Then,
\begin{equation}
\sum_\ell \ln(1+\lambda_{k,\ell}) = \ln \prod_\ell
(1+\lambda_{k,\ell})
=\ln f(-1)\,.
\end{equation}
The same can be done in the case $k<0$. Then, the grand potential can
be expressed as three integrals for $k$ in the domains
$\left]-\infty,0\right]$, $\left[0,k_-/2\right]$ and
$\left[k_-/2,+\infty\right[$. After a change of variable in the
integrals, the grand potential can be written in a more compact way as
\begin{eqnarray}
\label{eq:omega2murs}
\beta\omega=-\frac{k_0}{2\pi}
\big(\!\!
\int_{-k_-/2k_0}^{+\infty}\!\!\!+
\int_{-k_+/2k_0}^{+\infty}
\!\!\!\!\big)
\ln\bigg[
e^{-k_0h(\nu+1)}
\lefteqn{%
\moko
\bigg\{
\textstyle
I_{\nu}(\moko)K_{\nu+1}(\moko e^{-k_0h})}
\\
&
\textstyle
+
I_{\nu+1}(\moko e^{-k_0h})K_{\nu}(\moko)
\bigg\}\bigg]\,d\nu\,.
\nonumber
\end{eqnarray}

From now on we consider the simpler case when the container has infinite
width~($h\to\infty$). Equation~(\ref{eq:omega2murs}) becomes
\begin{equation}
\label{eq:omega1mur}
\beta\omega=
-\frac{k_0}{2\pi}
\big(
\int_{-k_-/2k_0}^{+\infty}+
\int_{-k_+/2k_0}^{+\infty}
\big)
\ln\left[
\left(\frac{2k_0}{m_0}\right)^{\nu}
\Gamma(\nu+1)I_\nu\left(\moko\right)
\right]\,d\nu
\,.
\end{equation}

The two-component plasma at $\beta q^2=2$ has logarithmic
divergences due to the collapse of pairs of particles of opposite
sign. This can be avoided by introducing hard core particles of radius
$R$.  The integrals in~(\ref{eq:omega2murs}) and~(\ref{eq:omega1mur})
are divergent and must be cutoff, the upper limit of integration
becomes $\nu_{\max}=(k_0 R)^{-1}$.

We can have an asymptotic expression for the grand potential in the
usual physical case where $k_0 \ll m_0$, using the Debye
expansion\citenote{Abra} of the Bessel functions. The calculations are
very similar to those of Ref.~\cite{JanManPis} for a different problem.
One finds
\begin{equation}
\omega=-\frac{1}{2k_0}p_0
+o\left(\moko\right)
\,,
\end{equation}
where $p_0$ is the pressure of a two-component plasma without an
external field. It is interesting to notice that this is the same
grand potential of a system without gravity, but confined in a
container of height $(2k_0)^{-1}$. Furthermore, from Ref.~\cite{tcp+g},
we know that most of the particles are in fact in the region
$0<y<(2k_0)^{-1}$ since for intermediate altitudes (the neutral zone)
the density profiles decay as $\exp(-2k_0 y)$.

\section{A special sum rule}

We can write a sum rule for this system by computing in two different
ways the force exerted on the bottom of the container. This force is
the weight of the column of fluid over the base of the container.
On the other hand, it is also given by the pressure at zero altitude
which can be computed by means of the contact theorem: it is the
density at zero altitude times $k_B T$. This gives the sum rule
\begin{equation}\label{sum-rule-long}
\rho_+(0)+\rho_-(0)=
\beta g\left(
M_+\int_0^{+\infty} \rho_+(y)\,dy
+
M_-\int_0^{+\infty} \rho_-(y)\,dy
\right)\,,
\end{equation}
where $\rho_\pm$ are the individual densities of the positive and
negative particles respectly.  This sum rule can be written in a
simpler way; because the system is neutral, both integrals
in~(\ref{sum-rule-long}) are equal to $n/2$, where $n$ is the total
number of particles. Then, the sum rule becomes
\begin{equation}\label{sum-rule}
\rho_+(0)+\rho_-(0)=k_0 n\,.
\end{equation}

We expect this to be true at any temperature. We can test the sum rule
in the present case when $\beta q^2=2$. The total number of particles
can be computed from the grand potential:
\begin{equation}
n=-\beta m_0\frac{\partial\omega}{\partial m_0}\,.
\end{equation}
From equation~(\ref{eq:omega1mur}) one finds
\begin{equation}
\label{eq:Ntot}
n=\frac{m_0}{2\pi}
\big(
\int_{-k_-/2k_0}^{+\infty}+
\int_{-k_+/2k_0}^{+\infty}
\big)
\frac{I_{\nu+1}(\moko)}{I_\nu(\moko)}
\,d\nu
\,.
\end{equation}
On the other hand, from equation~(3.10) of Ref.~\cite{tcp+g} we can
compute the densities at zero altitude:
\begin{equation}
\rho_\pm(0)=
\frac{m_0 k_0}{2\pi}
\int_{-k_\pm/2k_0}^{+\infty}
\frac{I_{\nu+1}(\moko)}{I_\nu(\moko)}
\,d\nu\,.
\end{equation} 
The sum rule~(\ref{sum-rule}) is verified.

\section*{Acknowledgment}

	I wish to thank B.~Jancovici for his remarks on an early
version of the manuscript.

\end{document}